\documentclass{emulateapj} 

\slugcomment{Submitted to ApJL}

\shorttitle{Where to find RIAFs}
\shortauthors{Chiaberge, M.}

\begin{document}

\title{Where to look for radiatively inefficient accretion flows (and find them)}


\author{M. Chiaberge\altaffilmark{1}}
\affil{Space Telescope Science Institute, 3700 San Martin Dr., Baltimore, MD 21218}
\email{chiab@stsci.edu}


\altaffiltext{1}{On leave from INAF, Istituto di Radioastronomia, via P. Gobetti 101, I-40129 Bologna}


\begin{abstract}

We have studied the nuclear emission detected in high-resolution radio
and optical data  of carefully selected samples of  low luminosity AGN
(LLAGN) in  the local universe.   When the Eddington ratio  is plotted
against  the  nuclear  ``radio-loudness''  parameter,  sources  divide
according  to  their physical  properties.   It  is  thus possible  to
disentangle  between nuclear  jets  and accretion  disks of  different
radiative efficiencies.  If this  simple interpretation is correct, we
now  have a powerful  tool to  investigate the  nature of  the nuclear
radiation, and identify radiatively inefficient accretion flows (RIAF)
candidates.  Our  results show that  the best chance  of investigating
RIAF processes in  the IR-to-UV spectral region is  to observe (at the
resolution provided by  HST) the nuclei of unobscured  Seyferts of the
lowest luminosity,  as well  as a sub-class  of LINERs.  In  all other
objects other radiation  processes dominate. In a sample  of 132 LLAGN
we identify  8 objects in which  we predict the radiation  from a RIAF
can be directly detected.

\end{abstract}

\keywords{galaxies: active --- galaxies: nuclei --- galaxies: jets ---
accretion, accretion disks}

\section{Introduction}

Models  of   radiatively  inefficient  accretion   flows  (RIAF)  were
originally developed  to reproduce the low state  of activity observed
in Cyg  X-1 \citep{ichimaru77}. They were later  discussed in relation
to AGN \citep{rees82,fabianrees95} in order to account for the lack of
activity  not  even  at   the  levels  predicted  by  Bondi  accretion
\citep{bondi52}.  The models are based  on the possibility that at low
accretion rates  a physical solution  for the accreting  flow involves
low  density,  hot, optically  thin  gas,  which  is also  radiatively
inefficient.   This regime would  occur instead  of that  of optically
thick, geometrically  thin disk accretion  (`Shakura--Sunyaev disks').
While  the theoretical effort  to characterize  accretion at  very low
rates   has    been   impressive   \citep[and    many   papers   since
then]{narayanyi},   direct observations  of RIAFs
still lack behind.

Low luminosity AGN  (LLAGN) are the class of  extragalactic objects in
which RIAFs  are most likely  to be at  work.  They show some  level of
activity  (compact radio emission  with high  brightness temperatures,
line emission,  nuclear X-ray emission)  which in the majority  of the
objects cannot be  explained in terms of processes  other than a faint
active  nucleus  powered  by  the  central  supermassive  black  hole.
However, because of their  intrinsic low luminosities, it is extremely
difficult to  study the emission from  RIAFs in LLAGN.   This holds in
particular  for  the wavelengths  at  which  the spectral  differences
between RIAF's and ``standard''  accretion disk models are expected to
differ the most, i.e. the IR-to-UV spectral region.  RIAFs should lack
both the "big blue-bump" and  the IR (reprocessed) bump, which instead
characterize  optically  thick,   geometrically  thin  accretion  disk
emission and the surrounding heated dust \citep[e.g][]{elvis94}.

While it is now possible to routinely disentangle the nuclear emission
from      that      of      the      host      thanks      to      HST
\citep[e.g.][]{hopeng,pap1,papllagn,sandrobarbara05},     the     RIAF
emission cannot be seen when  it is swamped by other nuclear radiation
processes, or  obscured by dust.  M~87  \citep{dimatteo03} is a
clear example of  that, where the nuclear emission  at all wavelengths
is  dominated by  non-thermal  synchrotron radiation.   \citet{Maoz05}
have  shown  that  among  a  sample  of 17  LLAGN,  15  of  them  show
variability over a timescale of  a few months, which demonstrate their
non--stellar origin.   However, it was  still unclear from  their work
whether  the nuclear radiation  is from  a jet  or from  the accretion
flow.

Recently,   among  LLAGN   which  show   very  low   Eddington  ratios
$L_{bol}/L_{Edd} << 10^{-3}$, \citet{papllagn} have found that a class
of  LLAGN,  mainly composed  by  LINERs  and low-luminosity  Seyfert~1
galaxies, show faint unresolved optical  nuclei in HST images that may
be  interpreted  as  direct  radiation  from  a  very  low  efficiency
accretion flow.  In fact, that  work shows that when the radio-optical
properties of LLAGN are considered, Seyfert, LINERs and low luminosity
(FR~I) radio  galaxies separate  into different regions  of diagnostic
planes,  according  to  the  properties  of  their  nuclei.   If  this
interpretation is correct, we now have a powerful tool to identify the
nature   of   the    nuclear   radiation   (i.e.    jet-dominated   or
accretion-dominated).   This picture  appears to  be confirmed  by the
nuclear SED of the low luminosity Seyfert galaxy NGC~4565, which shows
very  low obscuration  and  lacks the  typical  signature of  standard
accretion \citep{pap4565}.

In this Letter, by expanding the  sample of LLAGN, we discuss what is,
in our view, the most efficient way of identifying a sample of objects
that are  most likely to host  a RIAF and in  which IR-to-UV radiation
from the accretion process can be directly studied.

\section{Is the ``fundamental plane of black hole activity?'' 
a diagnostic plane?}

LLAGN have been  found to lie on the  so-called ``fundamental plane of
black hole activity''  \citep{merloni03,falcke04}, which has the great
merit of attempting  a unification of all sources  associated to black
holes, over  a large range  of masses and luminosities,  from Galactic
sources to powerful  quasars.  But the origin of  such a ``fundamental
plane''  and its  relationship  with the  physical  properties of  the
source is still a matter of debate \cite[e.g][]{bregman05,koerding06}.
Here  we want to  test whether  such a  plane can  be easily  used for
discriminating  between different  radiation  processes associated  to
inflows and/or  outflows around black  holes, with the aim  of finding
RIAF candidates.

\begin{figure}
\epsscale{1.2} \plotone{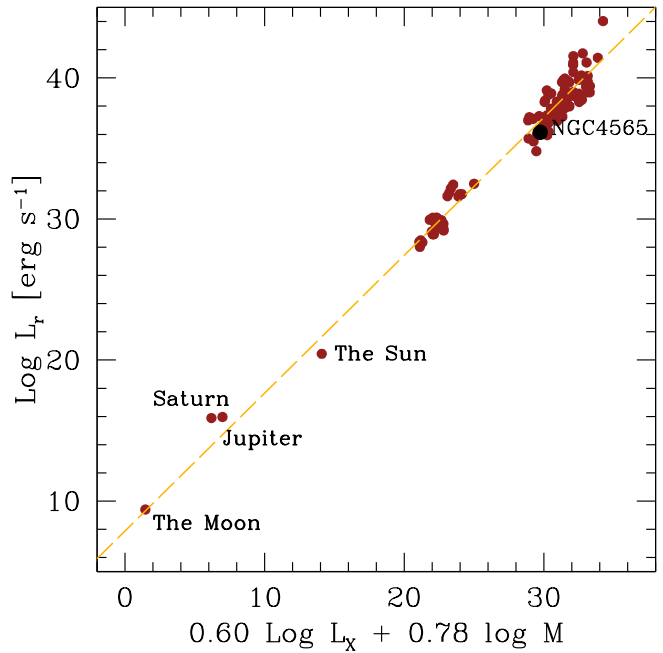}
\caption{The so-called  fundamental plane  of black hole  activity, to
which   we   over-plotted    solar   system   sources   \citep[adapted
from][]{merloni03}.  The location of  NGC~4565 (which  we refer  to in
Sect.~\ref{fund}) along the correlation is also shown.}
\label{funda}
\end{figure}

We  perform a  simple test,  to check  whether objects  with different
emission  processes and of  different physical  origin are  located in
distinct regions.  The ``fundamental plane'' is based on the radio and
X-ray  luminosity, combined  with the  black hole  mass, which  is the
ultimate ``energy source'' for the radiation we observe in the form of
inflows (accretion)  or outflows (jets).   However, if we  check where
other sources  of radio  and X-rays that  are not associated  to black
holes are  located in  this plane, we  find something  intriguing.  We
take the radio and X-ray fluxes of solar system objects, together with
their masses,  and we over plot  them onto the  ``fundamental plane of
black hole  activity''.  Unexpectedly, the Sun, the  Moon, Jupiter and
Saturn do fall on the same  plane, although they are tens of orders of
magnitudes fainter  and less massive  than all ``black  hole'' sources
(Fig.~\ref{funda}).  In  fact, the radiation  processes that originate
the emission from  the solar system objects are  variegated.  While it
is well  known that we observe  thermal radio emission  from the quiet
Sun,  Saturn,  the  Moon  and  Jupiter (below  3cm),  scattered  solar
radiation  is  the origin  for  the X-ray  emission  of  the Moon  and
probably Saturn, high ionization  emission lines dominate in the X-ray
Sun                and                Jupiter               \citep[see
e.g.][]{manson77,kraus86,schmitt91,ness04,branduardi06}.

This implies  that such  a ``fundamental plane''  might be  telling us
something profound  associated with the emitting power  of all sources
of  radio  and  X-ray  radiation,  but  cannot be  easily  used  as  a
diagnostic to  discriminate between different  radiation processes. On
the other  hand, the  apparent correlation might  only be a  result of
mere plotting luminosity  vs. luminosity, and ``artificially'' rescale
the quantities  with the mass  \citep{bregman05}. We speculate  that a
closer look at  the scatter of the ``correlation''  might lead to more
physical  information than just  trying to  reproduce it  assuming ``a
priori''  different models.   However,  as we  show  in the  following
section, in  order to discriminate different  radiation processes (and
eventually  find  RIAF  candidates)  we should  use  diagnostics  that
enhance the differences.

\section{How to find RIAFs: nuclear radio loudness and Eddington ratio}
\label{fund}

In order  to explore the nuclear  properties of all kinds  of LLAGN in
the local universe  with the aim of searching  for RIAF candidates, we
consider the following samples:

{\bf 1)} 33  FR~I Radio Galaxies (i.e. low  luminosity radio galaxies)
from the 3CR catalog (radio selected) \citep{pap1};

{\bf 2)} 25   Seyferts from the Palomar Survey  of nearby galaxies
and from the CfA sample, clearly limiting ourselves to those belonging
to Type~1, i.e.  unobscured, \citep[optically selected][]{hopeng};

{\bf 3)}  a complete sample  of 21 LINERs  from the Palomar  Survey of
nearby galaxies \citep[optically selected][]{ho97};

{\bf 4)} 51  nearby early-type galaxies (E+S0) with  radio emission $>
1$ mJy  at 5 GHz  (optical $+$ radio selection)  \citep[and references
therein]{sandrobarbara05}. The  large majority of the  galaxies in the
sample are spectrally classified as either LINER or Seyfert.

{\bf 5)} the 12 broad-line radio galaxies with  $z<0.3$ included the 3CR
catalog \citep[and references therein]{pap4}.

Samples 1, 2  and 3 have been examined  in detail in \citep{papllagn}.
Note  that  being  selected  according to  different  criteria,  these
objects do not constitute  a complete sample.  However, they perfectly
represent the  overall properties  of low power  active nuclei  in the
local universe, where RIAFs are most likely to be found. The sample of
nearby  ellipticals   partially  overlaps  with  samples   1)  2)  and
3). However, there are only 10  objects in common, so the total number
of objects considered is 132.

In \citet{papllagn} we showed that the nuclear properties of LLAGN are
best  understood  when  the   ratio  between  optical  luminosity  and
Eddington  luminosity $L_o/L_{Edd}$ is  plotted against  the ``nuclear
radio-loudness parameter'' $R$, defined  as the ratio between the {\it
nuclear} radio (core) luminosity at 5GHz $(F_{5GHz})$ as measured from
high-resolution (VLA or higher  resolution) data and the {\it nuclear}
optical luminosity  $L_o$ as measured  from HST images.  The  power of
this  diagnostic plane  resides  in the  fact  that we  only need  two
measured  quantities, plus  an estimate  of  the black  hole mass,  to
clearly separate sources of  different physical origin. This allows us
to  discriminate  between  jets   and  accretion  disks  of  different
radiative efficiency (and/or different accretion rates).

\begin{figure}
\epsscale{1.2} \plotone{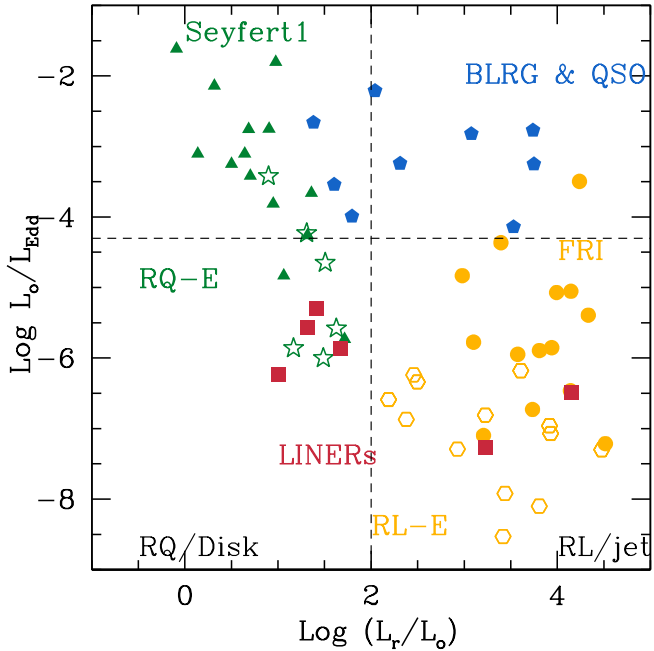}
\caption{The diagnostic plane for  the nuclear emission of LLAGN. The
ratio  between  the  optical  nuclear  luminosity  and  the  Eddington
luminosity is plotted against  the radio loudness parameters.  Symbols
are as follows: Seyfert~1 are  marked as triangles, pentagons are BLRG
and QSO  from the  3CR catalog, are  FR~I radio galaxies,  squares are
LINERs,  hexagons  are  early-type  ``core-galaxies'' and  stars  are
early-type ``power-law''  galaxies. The dashed lines  are only plotted
to guide  the eye in  discriminating between different  ``species'' of
nuclei.}
\label{ledd}
\end{figure}

Since here  we are  looking for object  harboring RIAF  candidates, in
Fig~\ref{ledd} we  plot galaxies for which the  ``active nucleus'' has
been detected  both in the radio  and the optical (or  near IR) bands.
The lack of detection of  compact radio emission is considered as lack
of evidence for AGN activity. On the other hand, the lack of detection
of the optical nucleus would set a double upper limit in the plot and,
most  importantly,   the  location  of   the  data  point   along  the
radio-loudness axis would  remain undetermined.  Therefore, we discard
such objects.   We also only plot  the objects for which  a black hole
mass estimate  is available, either  through gas dynamics or  by using
the relation with the stellar velocity dispersion.

Sources  separate into  4  quadrants: Seyfert~1s  occupy the  top-left
quadrant, FR~I radio galaxies are in the bottom-right quadrant.  These
two classes  define the regions characterized  by ``radio-quiet'' high
radiative   efficiency  accretion   and   ``radio-loud''jet  emission,
associated to low radiative efficient accretion, respectively.  LINERs
split  into   two  sub-samples:  both   are  in  the  region   of  low
$L_o/L_{Edd}$, some  of them  being radio quiet  and some  radio loud.
Nuclei  of ellipticals  separate according  to the  properties  of the
radial  brightness profile  of the  host  \citep{sandrobarbara06}, and
behave as the LINERs: core-galaxies are radio-loud, power-law galaxies
are  radio-quiet  \citep[for  a   definition  of  core  and  power-law
galaxies, see  e.g.][]{faber97}.  Broad line  radio galaxies, instead,
are found  at the top  of the plot,  at high values  of $L_o/L_{Edd}$,
while their location on the radio loudness axis most likely depends on
the relative  importance of the jet  and disk emission  in the optical
band  (the radio  core being  dominated  by radiation  from the  jet).
Although it is not easy  to identify any ``bi-modality'' when plotting
all the  samples together as  done in Fig.~\ref{ledd}, we  draw dashed
lines  just  to  guide  the  eye in  discriminating  between  physical
processes  that are  clearly different.   Clearly, there  is a  bit of
ambiguity for the nuclei that fall close to those ``dividing lines''.

Let us  now focus on the  right hand side  of the panel. Here  we find
jet-dominated  nuclei  (LLRG and  a  subsample  of  LINERs). In  these
objects,  both the  radio and  the  optical radiation  is most  likely
synchrotron  emission  from  the base  of  the  jet,  as it  has  been
established      for      low      luminosity      radio      galaxies
\citep[e.g.][]{pap1,gijs}.  The  radiation from the  accretion flow is
swamped by the jet emission,  and cannot be studied directly (at least
in the observing bands considered  here).  We should point out that in
this   case   we   have   an   ``upper  limit''   to   the   Eddington
ratio. Nevertheless, for  several objects, this upper limit  is as low
as $L_o/L_{Edd} \sim 10^{-8}$. On the other hand, none of the detected
objects on the left side of  the plot reaches such low values.  On the
other hand,  these extremely low  efficient accreting black  holes are
still capable of producing a  ``jet''.  This means that although RIAFs
(or whatever  these objects are!)   are inefficient from the  point of
view of producing radiation, under certain circumstances (i.e. for the
``radio-loud''  nuclei)  they  can  be  very  efficient  in  producing
outflows  or  even  (relativistic)  jets,  as  in  the  case  of  LLRG
\citep[see e.g.][]{rees82}.   In \citet{papllagn} we  pointed out that
``radio-loud'' (i.e.  jet dominated) nuclei tend to be associated with
the most massive  black holes.  Thus it is  tempting to speculate that
the enhanced efficiency in  producing powerful jets is somehow related
to the  higher black hole mass.  An alternative picture is  that it is
the   black    hole   spin   that   efficiently    powers   the   jets
\citep{blandfordsaasfee,sikora06},  or possibly  a combination  of the
two proposed scenarios.

Let us now  focus on the left  side of the plot, which  is central for
the   purposes  of   this   Letter.   As   discussed   in  detail   in
\citet{papllagn}, in that region we  find nuclei that show a low value
of  $L_r/L_o$, i.e.   an optical  excess with  respect to  the optical
counterpart  of the  radio (synchrotron)  radiation.   For Seyfert~1s,
such an  optical excess  is readily interpreted  as emission  from the
accretion disk.  A  few bright Seyferts for which  the nuclear SED has
been     derived     support     this    interpretation     \citep[see
e.g.][]{alonso03,pap4565}.   For the brightest  nuclei, the  limits on
the Eddington  ratio are  still compatible with  radiatively efficient
accretion.   Furthermore, we should  stress that  for those  objects a
bolometric correction of  a factor $\sim 10$ \citep{elvis94,marconi04}
should be  performed, because we know  their SED and we  know that the
optical $R$ or  $V$ band do not correspond to  the peak of radiatively
efficient accretion energy output.  On  the other hand, for objects of
the  lowest  Eddington  ratios,  we  don't know  what  the  bolometric
correction should  be, but  a factor of  10 (or even  slightly higher)
would not  change our conclusions.   Objects with Eddington  ratios as
low  as  $10^{-6}$  cannot  be  anyhow  reconciled  with  ``standard''
accretion  disk models.  Therefore,  they are  the best  candidates to
host radiatively inefficient accretion disks.

In  fact,  we have  shown  in  \citet{pap4565}  that NGC~4565,  a  low
luminosity  Seyfert (not  included in  the sample  considered  in this
Letter) that resides in the lower-left region of the diagnostic plane,
shows an unusual SED which  does not display the typical signatures of
radiatively efficient accretion disks (i.e.  lacks both the UV and the
IR bumps).  This strongly supports our prediction that objects in that
part of the plot are not ``standard'' disks.

Since we want  to identify the objects in which  radiation from a RIAF
can be detected and studied  in the IR-to-UV spectral region, in order
to derive  the SED and  set constraints to  the models, our  best RIAF
candidates are  those in which a  nuclear source has  been detected in
archival  HST images.   This reduces  the number  of objects  to eight
candidates, plus  NGC~4565 which  has been already  studied elsewhere.
Summarizing, we predict that RIAFs  can be detected and studied in the
following  objects:  M~81,  NGC~3245,  NGC~3414,  NGC~3718,  NGC~3998,
NGC~4143, NGC~4203, NGC~4565 and NGC~4736.

\section{Summary and conclusions}

By combining the Eddington  ratio as measured from detected unresolved
nuclei in HST images  and the nuclear ``radio-loudness'' parameter, we
have found  a straightforward way of  distinguishing radiation emitted
by  the   accretion  process  from   jets  in  LLAGN  and   find  RIAF
candidates. At the end of  the selection process, which started from a
large sample of 132 AGN in  the local universe, we are left with eight
RIAF candidates for  which the nuclear source is  detected in archival
HST images. These are the best targets for follow-up studies, not only
in  the IR-to-UV  spectral region,  but also  in other  bands  such as
far-IR, mm, and  X-rays. Studying those objects in  more detail is the
best  chance to provide  constraints for  RIAF models  associated with
supermassive black holes.

The method to find RIAF candidates outlined in this Letter is based on
the radio and optical emission.  However, it is possible to extend our
ability of discovering  new candidates using other bands  in which the
nuclear luminosity can be estimated. For example, nuclear IR radiation
between $\sim 10$ and $\sim 20 \mu$m can be isolated in nearby objects
using  8m  class  telescopes,  as done  by  \citet{whysong}.   Another
possible way  of estimating  the nuclear flux  is to use  the emission
line strength, assuming a correlation between these two quantities, as
\citet[e.g.][]{sandrobarbara06}  pointed  out  that  when  the  [OIII]
luminosity is used instead of  the optical nuclear luminosity from HST
images,  similar   results  are  achieved.   However,   in  all  these
alternative methods  we are  not sure that  the optical source  can be
detected in order to derive the SED and constrain the models.

We should also  stress that the diagnostic plane  discussed here holds
for LLAGN  in the local universe.   When high-z more  powerful AGN are
over-plotted,  cosmic evolution  may play  a role  in  determining the
position of  the nuclei  in the  plane. For example,  since it  is not
clear  whether the  local  relations between  properties  of the  host
galaxy and the supermassive black  hole mass can be easily extended to
high-redshift galaxies, the estimate of the black hole mass may change
substantially  with cosmic  time.  Furthermore,  the  overall spectral
properties of jets and disks, and their relative importance in the SED
may  change as  well, and  all this  should be  carefully investigated
before including ``all known AGNs'' in the plane and draw conclusions.
However, RIAFs are  so faint and so difficult to  detect, that at this
stage it is best to look for them in local objects.

\acknowledgments M.C.   wishes to thank  A.  Capetti, A.   Celotti, D.
Macchetto,  R.   Gilli and  W.B.  Sparks  for  useful discussions  and
insightful comments.






\begin{thebibliography}{32}
\expandafter\ifx\csname natexlab\endcsname\relax\def\natexlab#1{#1}\fi

\bibitem[{{Alonso-Herrero} {et~al.}(2003){Alonso-Herrero}, {Quillen}, {Rieke},
  {Ivanov}, \& {Efstathiou}}]{alonso03}
{Alonso-Herrero}, A., {Quillen}, A.~C., {Rieke}, G.~H., {Ivanov}, V.~D., \&
  {Efstathiou}, A. 2003, \aj, 126, 81

\bibitem[{{Blandford} {et~al.}(1990){Blandford}, {Netzer}, {Woltjer},
  {Courvoisier}, \& {Mayor}}]{blandfordsaasfee}
{Blandford}, R.~D., {Netzer}, H., {Woltjer}, L., {Courvoisier}, T.~J.-L., \&
  {Mayor}, M. 1990, {Active Galactic Nuclei} (Saas-Fee Advanced Course
  20.~Lecture Notes 1990.~Swiss Society for Astrophysics and Astronomy, XII,
  280 pp.~97 figs..~ Springer-Verlag Berlin Heidelberg New York)

\bibitem[{{Bondi}(1952)}]{bondi52}
{Bondi}, H. 1952, \mnras, 112, 195

\bibitem[{{Branduardi-Raymont} {et~al.}(2006){Branduardi-Raymont}, {Bhardwaj},
  {Elsner}, {Gladstone}, {Ramsay}, {Rodriguez}, {Soria}, {Waite}, \&
  {Cravens}}]{branduardi06}
{Branduardi-Raymont}, G., et al. 2006, in ESA SP-604: The X-ray Universe 2005, ed.
  A.~{Wilson}, 15--20

\bibitem[{{Bregman}(2005)}]{bregman05}
{Bregman}, J.~N. 2005, ArXiv Astrophysics e-prints

\bibitem[{{Capetti} \& {Balmaverde}(2005)}]{sandrobarbara05}
{Capetti}, A., \& {Balmaverde}, B. 2005, \aap, 440, 73

\bibitem[{{Capetti} \& {Balmaverde}(2006)}]{sandrobarbara06}
---. 2006, \aap, 453, 27

\bibitem[{{Chiaberge} {et~al.}(1999){Chiaberge}, {Capetti}, \&
  {Celotti}}]{pap1}
{Chiaberge}, M., {Capetti}, A., \& {Celotti}, A. 1999, \aap, 349, 77

\bibitem[{{Chiaberge} {et~al.}(2002){Chiaberge}, {Capetti}, \&
  {Celotti}}]{pap4}
---. 2002, \aap, 394, 791

\bibitem[{{Chiaberge} {et~al.}(2005){Chiaberge}, {Capetti}, \&
  {Macchetto}}]{papllagn}
{Chiaberge}, M., {Capetti}, A., \& {Macchetto}, F.~D. 2005, \apj, 625, 716

\bibitem[{{Chiaberge} {et~al.}(2006){Chiaberge}, {Gilli}, {Macchetto}, \&
  {Sparks}}]{pap4565}
{Chiaberge}, M., {Gilli}, R., {Macchetto}, F.~D., \& {Sparks}, W.~B. 2006,
  \apj, 651, 728

\bibitem[{{Di Matteo} {et~al.}(2003){Di Matteo}, {Allen}, {Fabian}, {Wilson},
  \& {Young}}]{dimatteo03}
{Di Matteo}, T., {Allen}, S.~W., {Fabian}, A.~C., {Wilson}, A.~S., \& {Young},
  A.~J. 2003, \apj, 582, 133

\bibitem[{{Elvis} {et~al.}(1994){Elvis}, {Wilkes}, {McDowell}, {Green},
  {Bechtold}, {Willner}, {Oey}, {Polomski}, \& {Cutri}}]{elvis94}
{Elvis}, M., {Wilkes}, B.~J., {McDowell}, J.~C., {Green}, R.~F., {Bechtold},
  J., {Willner}, S.~P., {Oey}, M.~S., {Polomski}, E., \& {Cutri}, R. 1994,
  \apjs, 95, 1

\bibitem[{{Faber} {et~al.}(1997){Faber}, {Tremaine}, {Ajhar}, {Byun},
  {Dressler}, {Gebhardt}, {Grillmair}, {Kormendy}, {Lauer}, \&
  {Richstone}}]{faber97}
{Faber}, S.~M., et al. 1997, \aj, 114, 1771

\bibitem[{{Fabian} \& {Rees}(1995)}]{fabianrees95}
{Fabian}, A.~C., \& {Rees}, M.~J. 1995, \mnras, 277, L55

\bibitem[{{Falcke} {et~al.}(2004){Falcke}, {K{\"o}rding}, \&
  {Markoff}}]{falcke04}
{Falcke}, H., {K{\"o}rding}, E., \& {Markoff}, S. 2004, \aap, 414, 895

\bibitem[{{Ho} {et~al.}(1997){Ho}, {Filippenko}, \& {Sargent}}]{ho97}
{Ho}, L.~C., {Filippenko}, A.~V., \& {Sargent}, W.~L.~W. 1997, \apjs, 112, 315

\bibitem[{{Ho} \& {Peng}(2001)}]{hopeng}
{Ho}, L.~C., \& {Peng}, C.~Y. 2001, \apj, 555, 650

\bibitem[{{Ichimaru}(1977)}]{ichimaru77}
{Ichimaru}, S. 1977, \apj, 214, 840

\bibitem[{{Koerding} {et~al.}(2006){Koerding}, {Falcke}, \&
  {Corbel}}]{koerding06}
{Koerding}, E., {Falcke}, H., \& {Corbel}, S. 2006, ArXiv Astrophysics e-prints

\bibitem[{{Kraus}(1986)}]{kraus86}
{Kraus}, J.~D. 1986, {Radio astronomy} (Powell, Ohio: Cygnus-Quasar Books,
  1986)

\bibitem[{{Manson}(1977)}]{manson77}
{Manson}, J.~E. 1977, in The Solar Output and its Variation, ed. O.~R. {White},
  261--+

\bibitem[{{Maoz} {et~al.}(2005){Maoz}, {Nagar}, {Falcke}, \& {Wilson}}]{Maoz05}
{Maoz}, D., {Nagar}, N.~M., {Falcke}, H., \& {Wilson}, A.~S. 2005, \apj, 625,
  699

\bibitem[{{Marconi} {et~al.}(2004){Marconi}, {Risaliti}, {Gilli}, {Hunt},
  {Maiolino}, \& {Salvati}}]{marconi04}
{Marconi}, A., {Risaliti}, G., {Gilli}, R., {Hunt}, L.~K., {Maiolino}, R., \&
  {Salvati}, M. 2004, \mnras, 351, 169

\bibitem[{{Merloni} {et~al.}(2003){Merloni}, {Heinz}, \& {di
  Matteo}}]{merloni03}
{Merloni}, A., {Heinz}, S., \& {di Matteo}, T. 2003, \mnras, 345, 1057

\bibitem[{{Narayan} \& {Yi}(1995)}]{narayanyi}
{Narayan}, R., \& {Yi}, I. 1995, \apj, 444, 231

\bibitem[{{Ness} {et~al.}(2004){Ness}, {Schmitt}, {Wolk}, {Dennerl}, \&
  {Burwitz}}]{ness04}
{Ness}, J.-U., {Schmitt}, J.~H.~M.~M., {Wolk}, S.~J., {Dennerl}, K., \&
  {Burwitz}, V. 2004, \aap, 418, 337

\bibitem[{{Rees} {et~al.}(1982){Rees}, {Phinney}, {Begelman}, \&
  {Blandford}}]{rees82}
{Rees}, M.~J., {Phinney}, E.~S., {Begelman}, M.~C., \& {Blandford}, R.~D. 1982,
  \nat, 295, 17

\bibitem[{{Schmitt} {et~al.}(1991){Schmitt}, {Snowden}, {Aschenbach},
  {Hasinger}, {Pfeffermann}, {Predehl}, \& {Trumper}}]{schmitt91}
{Schmitt}, J.~H.~M.~M., {Snowden}, S.~L., {Aschenbach}, B., {Hasinger}, G.,
  {Pfeffermann}, E., {Predehl}, P., \& {Trumper}, J. 1991, \nat, 349, 583

\bibitem[{{Sikora} {et~al.}(2006){Sikora}, {Stawarz}, \& {Lasota}}]{sikora06}
{Sikora}, M., {Stawarz}, L., \& {Lasota}, J.~. 2006, ArXiv Astrophysics
  e-prints

\bibitem[{{Verdoes Kleijn} {et~al.}(2002){Verdoes Kleijn}, {Baum}, {de Zeeuw},
  \& {O'Dea}}]{gijs}
{Verdoes Kleijn}, G.~A., {Baum}, S.~A., {de Zeeuw}, P.~T., \& {O'Dea}, C.~P.
  2002, \aj, 123, 1334

\bibitem[{{Whysong} \& {Antonucci}(2004)}]{whysong}
{Whysong}, D., \& {Antonucci}, R. 2004, \apj, 602, 116

\end{thebibliography}

\end{document}